%% file: main.tex
\documentclass[5p,times]{elsarticle}

\usepackage{amssymb}

\usepackage{hyperref} 
\usepackage{xurl} 
\usepackage{adjustbox} 
\usepackage{svg} 
\usepackage{wasysym} 
\journal{Future Generation Computer Systems}

\begin{document}

\begin{frontmatter}



\title{A Survey on Heterogeneous Computing Using SmartNICs and Emerging Data Processing Units}

\author[MST]{Nathan Tibbetts\corref{corresponding_auth}}
\ead{nathan.tibbetts@mst.edu}
\author[MST]{Sifat Ibtisum\fnref{LAMAR}}
\ead{sibtisum@lamar.edu}
\author[MST]{Satish Puri}
\ead{satish.puri@mst.edu}
\cortext[corresponding_auth]{Corresponding author.}
\fntext[LAMAR]{Present address: Lamar University, Department of Computer Science 211 Red Bird Lane, BOX 10056, Beaumont, Texas, 77710, USA.}

\affiliation[MST]{organization={Department of Computer Science, Missouri University of Science and Technology},
            addressline={500 West 15th Street, 325 Computer Science Bldg.}, 
            city={Rolla},
            postcode={65409}, 
            state={Missouri},
            country={USA}}



\begin{abstract}
    The emergence of new, off-path smart network cards (SmartNICs), known generally as Data Processing Units (DPU), has opened a wide range of research opportunities. Of particular interest is the use of these and related devices in tandem with their host's CPU, creating a heterogeneous computing system with new properties and strengths to be explored, capable of accelerating a wide variety of workloads. This survey begins by providing the motivation and relevant background information for this new field, including its origins, a few current hardware offerings, major programming languages and frameworks for using them, and associated challenges. We then review and categorize a number of recent works in the field, covering a wide variety of studies, benchmarks, and application areas, such as data center infrastructure, commercial uses, and AI and ML acceleration. We conclude with a few observations.
\end{abstract}




\end{frontmatter}



\section{Introduction}
The computational hardware landscape is changing rapidly with the advent of new processors and networking equipment. In this survey, we review academic and industrial research using an emerging class of advanced network computing devices known as data processing units (DPU), with a particular focus on their use in heterogeneous and parallel computing applications, in contrast to more general surveys, such as the recent work of Kfoury et al.~\cite{comprehensive_expert_survey}. Such preexisting works are also more expert-oriented and lack the applications focus of ours; this survey is instead intended to be very accessible to those less familiar with networking devices, but who want to use them in applications.

As network interface cards (NIC) led to the development of smart network interface cards (SmartNIC), we see similar advancements in the field of SmartNICs. Recent developments in SmartNICs by companies like NVIDIA, Amazon, Microsoft (Fungible acquisition) and AMD (Pensando acquisition) have led to the new ``DPU'' nomenclature.

A DPU is thus essentially a class of advanced SmartNICs, with distinguishing aspects such as typically integrating a general-purpose CPU capable of running an operating system (OS), with DRAM and storage, into the high-speed NIC datapath of a SmartNIC. However, the CPU cores in a DPU are generally cheaper and less powerful than the host's CPU cores~\cite{increasingdatacenter}, and some DPUs incorporate powerful FPGAs in place of or in addition to CPUs. DPUs also have a high data throughput, often 100-400 Gbps. To provide an example, NVIDIA's Mellanox BlueField devices are considered to be DPUs, and their flagship DPU product is the BlueField-3; NVIDIA is one of the current industry leaders and a trend-setter in our line of research. Other examples are listed in Section~\ref{sec:dpumarket}. DPUs are typically used in place of a server's standard SmartNICs, which connect it to the network; they are an increasingly integral component of conventional data centers' infrastructure. For this survey's heterogeneous and parallel computing focus, CPU-based and OS-capable devices are of particular interest.

Note that while we use the term `DPU' as described, the nomenclature is in reality inconsistent and depends upon the vendor and product specialization. For example, Intel's ``Infrastructure Processing Unit'' (IPU) (not to be confused with an ``Intelligence Processing Unit'', which is a fundamentally different device) can be considered a DPU, despite its different name. Furthermore, the term `Data Processing Unit' itself has become overloaded, as it has also been used to describe other devices, such as Processing-in-Memory (PIM) devices, which are not discussed in this work, and the term ``DPU'' can refer to Deep-Learning Processing Units; neither of these are SmartNICs. In a rather different paradigm, some companies, such as NVIDIA, would argue that the real DPU is actually the specialized processing unit (different from a CPU or a GPU) often included in a SmartNIC, with features irreplaceable by a standard CPU~\cite{whatisadpu}. In our work, we generally adhere to the network card--oriented description of a DPU provided by ServeTheHome~\cite{21thsmartNIC}, discussed in Section~\ref{sec:evolution}, and use it as a reasonably general differentiation between DPUs and other SmartNICs.

The development of advanced SmartNICs, or DPUs, was started by large cloud service providers (``hyperscalers'') such as Amazon who realized that around 30\% of processor cores were being utilized for infrastructure processing~\cite{ubuntuSmartNIC2023}. This motivated cloud vendors to develop advanced SmartNICs to offload infrastructure-related tasks related to networking, security, and storage from host CPUs, so that host CPU cores can focus on users' application workloads. According to the projection of Dell'Oro Group, the traditional NIC market share will be overtaken by SmartNICs by 2027~\cite{delloro}. We discuss the motivation behind DPU creation and use in Section~\ref{sec:motivation}, and their evolution in~\ref{sec:evolution}.

There are many different reasons for developers and professionals to use SmartNICs, the most obvious being their intended use as network infrastructure support devices. For example, one such use case is to perform network offload, where networking tasks, such as packet processing, are delegated to SmartNICs by host CPUs. Their utility extends beyond basic networking functions and is also capable of encompassing security offloading, which is achieved by accelerating cryptographic operations such as those common in a Secure Sockets Layer (SSL), Transport Layer Security (TLS), and encryption. Similarly, SmartNICs are used to offload storage-related operations, as well as to support virtualization tasks.

However, data center infrastructure is not the sole application domain in which SmartNICs are applied, providing the main motivation for this paper. As the role of SmartNICs has grown to capably include offloading virtually any task, treating DPUs as heterogeneous compute nodes, rather than simply accelerators, this survey focuses on works exploring their use in a variety of application-level tasks, in addition to infrastructure acceleration. Such new offload methods might be described by terms such as in-network computing, edge computing, communication acceleration, or simply heterogeneous parallel computing. In both infrastructure and applications, such accelerations are made possible primarily by two things: 1) acceleration and offload of computations and networking functions, and 2) isolation between host and DPU. We discuss offloading briefly in Section~\ref{sec:acceleration}, how data is moved in~\ref{sec:data_path}, and isolation in~\ref{sec:isolation}. Section~\ref{sec:challenges} discusses challenges regarding their use, and Section~\ref{sec:dpumarket} presents the variety of DPU offerings by technology vendors.

After providing background information on the above topics, the remainder of the survey dives into the current research. Section~\ref{frameworks} provides an overview of various programming tools and frameworks for developing application software targeted to DPUs, Section~\ref{section_studies_and_benchmarks} cites a number of broader studies, benchmarking tools, and works which give a big-picture view, and Section~\ref{apparea} discusses various applications of DPUs, including topics such as big data, HPC, and machine learning. As DPUs continue to evolve, their application areas and benchmarks, and the focal points chosen by the research community, will play a pivotal role in shaping the landscape of high-performance computing and data processing. Finally, we conclude with our broad observations and recommendations regarding the field, and which research areas seem most effective, in Section~\ref{conclusion}.


\section{Background}
\label{background}

For those unfamiliar with SmartNICs, a certain amount of background information will help to understand the research we review. These background sections are intended to be a brief overview; other surveys, such as that by Kfoury et al.~\cite{comprehensive_expert_survey}, cover these topics in much greater depth for the benefit of domain experts.


\subsection{The Purpose of a DPU} 
\label{sec:motivation}

Approximately 30\% of the work done in a data center is in handling network infrastructure tasks (called an ``infrastructure tax''), which means that it requires almost half again as many servers to get the intended work done; this can be seen as severely bloating the size and cost of data centers. DPUs are increasingly seen as a solution to this problem, and CPUs as insufficient \cite{whatSmartNIC2023}.

Heterogeneous computing using emerging Data Processing Units (DPUs) is driven in particular by the necessity for specialized acceleration, enhanced performance, and energy efficiency. DPUs, which are specifically designed for particular types of tasks, optimize the effectiveness of the system by transferring specialized workloads from the host CPU, providing a more energy-efficient solution for data center infrastructure workloads, which also improves cost-efficiency. However, since DPUs have a general-use processor that is also more energy efficient~\cite{increasingdatacenter}, they are capable of accommodating a wide range of workloads and enhancing the overall utilization of resources. In this way, the adaptability and flexibility of heterogeneous computing with DPUs is intended to address the evolving requirements of modern applications~\cite{IPU-basedComputingPlatform}.

However, the motivation behind the creation of these smart network devices goes deeper. As the world turns to cloud technologies and flexibly hosted web services, there is an increasing need for software-defined networking (SDN)---systems capable of not only specialized, but customizable networking configurations within data centers and supercomputers~\cite{Caulfield2018}---perhaps for both infrastructure and tenant applications alike. DPUs will be part of the future infrastructure that powers services such as augmented reality, smart cities, edge computing, big-data analytics, and more~\cite{exploringtheadvantagesofdeployingdpus}.

A `smarter' NIC understandably comes with a higher price point and per-node power usage, and whether or not that cost comes out to be worth it is a matter of balance, requiring the consideration of many factors for a given data center, not least of which is the potential for millions of dollars worth of electricity and capital savings for a large data center \cite{increasingdatacenter}. However, the desire for software-reconfigurable optimization within the data center appears to generally be worth the cost, given the directions in which technology is advancing; DPU adoption is expected to keep increasing~\cite{DataCenterDynamics}.


\subsection{Evolution of NICs and DPUs}
\label{sec:evolution}

\begin{figure*}[h]
    \centering
    \includegraphics[width=380pt]{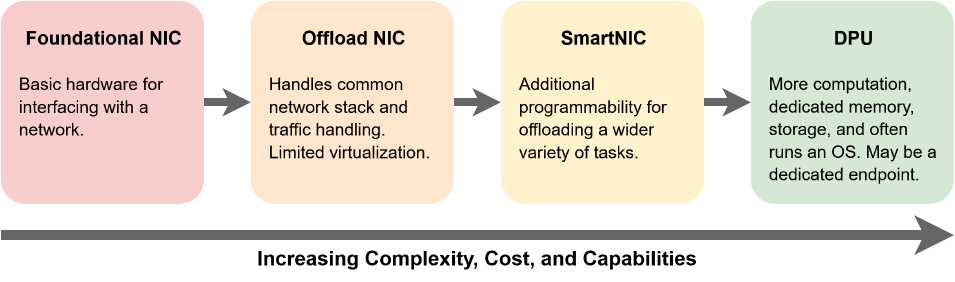}
    \caption{Part of the 2021 ServeTheHome NIC Continuum, a useful classification of existing NIC hardware, and which doubles as a representation of its evolution over time~\protect\cite{21thsmartNIC}.}
    \label{fig:nic-evolution}
\end{figure*}

The NIC class distinctions shown in Figure \ref{fig:nic-evolution} are not universally recognized, but nevertheless represent a useful categorization created by ServeTheHome (STH) \cite{21thsmartNIC}, which we will use. This progression evolved quite naturally over time, beginning with companies such as Amazon designing in-house solutions (namely AWS Nitro) to meet a growing need for more secure and efficient system management and virtualization services. These pioneers of the field represented a reimagining of network and data center infrastructure, and others followed who built upon these concepts in search of more generalizable solutions~\cite{STHQuickPrimer, AWSNitro}.

The `Foundational NIC' shown in Figure \ref{fig:nic-evolution} represents the fundamental level of a network interface. Nearly all contemporary NICs possess certain rudimentary offloads, such as IPv4, IPv6, and TCP/UDP checksum offloads. Foundational NICs are specifically engineered to facilitate low-cost network ports, eschewing many of the advanced offload features that contribute to increased cost of more sophisticated devices. Foundational NICs retain their significance within the industry, but are insufficient in many cases, particularly at higher data rates where the escalation in data processing necessitates enhanced computational capabilities.

Consequently, the prevailing NICs are predominantly mid-tier `Offload NICs', which generally provide networking speeds of about 100Gbps and above \cite{21thsmartNIC}. In order to design NICs with such speeds, it becomes necessary to offload more network support functions, generally in the form of on-board hardware accelerators (ASICs, or application-specific integrated circuits) \cite{21thsmartNIC}; otherwise, the host CPU would be quite busy trying to keep up.

`SmartNICs', however, go beyond simple offloads. These are NICs that possess programmable pipelines, allowing systems and applications to tweak and configure offloads more flexibly. The primary purpose of a SmartNIC, as being part of a server device, is to reduce the burden imposed on the host CPU \cite{21thsmartNIC}. ``Smart'', in the name, suggests a certain programmability, or at the very least an increased ability to handle complex tasks---both current, and being adaptable to future needs~\cite{composableDataServices2023}.

A `DPU' differs from this paradigm, however, in that it can hardly be considered part of the host server; it is, in and of itself, a mini network endpoint \cite{21thsmartNIC, dpuvssmartnic2023}, providing `off-path' capabilities (we will define off-path in Section~\ref{sec:data_path}). The concept of a DPU encompasses the functionality of a SmartNIC and extends beyond it, possessing not only ASIC offload capabilities and the flexible programmable pipelines of a SmartNIC, but also its own memory, storage, and CPUs or FPGAs. This design enables them not only to perform network tasks, but to participate in \textit{data} processing tasks---further freeing up the host's compute resources, all while consuming less power than the host \cite{dpuvssmartnic2023}. While ServeTheHome further differentiates DPUs from more ``Exotic'' devices incorporating large Field-Programmable Gate Arrays (FPGA), we lump these in with DPUs for the purposes of this survey. We note also NVIDIA's `SuperNIC' product line as an important type of SmartNIC; however, as these devices primarily focus on inter-GPU communication, rather than specializing in providing heterogeneous internal computation themselves, we do not discuss them further in this survey~\cite{whatisasupernic}.


\subsection{Computational Acceleration and Offload}
\label{sec:acceleration}

While NICs and SmartNICs already offload some types of work from the host's CPUs, DPUs are designed to go further, being able to offload virtually any kind of work, with generally the same goal of freeing up host resources to focus on application-level processing. A DPU, being `close to' the network, boasts better network I/O processing performance compared to the host, making it ideal for accelerating network stack and packet processing, session management, distributed storage and file system client execution, erasure coding, network security layers, or acting as a virtual switch \cite{dpfs2023, ProceedingswithsmartNIC, increasingdatacenter}. DPUs have also been leveraged in network-edge scenarios for applications such as network awareness, power-efficient edge-to-cloud continuum, etc. \cite{dpuedge2022, dpubench2023}. The goal of such offloads is to free up the host's resources to focus on applications processing tasks and improve system-wide performance by reducing network and data center overhead \cite{NovelFramework2023ieee}.

\begin{figure}[h]
    \centering
    \includegraphics[width=5.5cm]{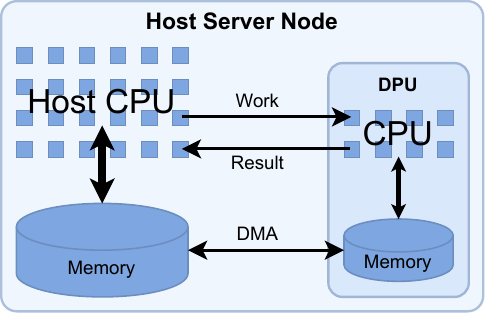}
    \caption{Illustration of offloading work to DPU hardware within a compute node. Offload methods can take advantage of a variety of tools, such as Direct Memory Access (DMA), illustrated here.}
    \label{ComputeNodeDiagram}
\end{figure}

And, while DPU acceleration typically refers to offloading these types of network and infrastructure tasks using on-board hardware accelerators (ASICs) \cite{dpusolution2022, dpubench2023}, this survey especially points out those works which instead show that one can offload generic computations and data manipulation to a DPU, including work in the applications layer, moving some processing from the host to the DPU's CPUs. Some notable examples from the papers we review include big data processes, machine learning functions and inference, molecular dynamics, and various other forms of scientific computing. A simple illustration of what offloading to a DPU might look like is provided in Figure~\ref{ComputeNodeDiagram}.

\subsection{Data Path}
\label{sec:data_path}

\input{Table_1}


When discussing `data paths' in terms of advanced SmartNICs such as DPUs, the term may be more specific than the typical usage. In general, a data path describes the steps through which data travels as it is moved, analyzed, transformed, and processed within a workflow. In a DPU, however, the data path usually traverses both the host and the DPU, and typically can be described one of two ways: `on-path' or `off-path', as shown in Table \ref{table:data_path}. Understanding the difference between the two is helpful in understanding the various use-cases of SmartNICs and DPUs. In this work, we generally associate the term DPU with off-path devices.

On-path SmartNICs have CPU cores that are situated on the communication path of the network packets traveling to the host, and possess the ability to manipulate each incoming or outgoing packet rapidly. Too much additional processing assigned to these cores can bottleneck the communication path, because all network traffic must pass through them. Examples of on-path SmartNICs include the LiquidIO II CN2350 and CN2360.

By contrast, off-path SmartNICs (typically DPUs) have cores that are not in the direct line of traffic to the host. Their internal NIC can use customizable network forwarding rules to sort between packets that should be sent to the DPU cores first for preprocessing, and those that should be forwarded directly to the host instead. However, direct control over packets and memory will be limited, and will have overheads; any off-path processing which occurs lengthens the data path, and thus time, for packets to make it to the host \cite{ComprehensiveStudyDPU2023}. Examples of off-path SmartNICs include the Mellanox BlueField and the Broadcom Stingray \cite{smartNICsiPipe2019}.

In short, a `DPU' is a class of device with certain advanced capabilities, which is usually an `off-path' device, a term which describes both the physical network path within the device, and a paradigm for how the data flows through it. Off-path processing might include, for example, live network flow analysis, data preprocessing steps, etc. Many research projects take advantage of the flexibility of off-path systems, analyzing the data path and data flow of a given application and experimenting with using a SmartNIC to optimize its efficiency, or to save on data preprocessing steps. Others benchmark the DPUs themselves in different path modes~\cite{kashyap_idiosyncrasies}.


\subsection{Isolation}
\label{sec:isolation}

By DPU isolation, we are actually referring to two separate concepts, performance isolation and security isolation, both of which make these devices valuable in data center ecosystems.

First, amid the many services or tenants which a server node often has to support, being able to separate or isolate many of the infrastructure tasks from the tenant or application workload, as well as separating those tenant workloads from each other, has the capacity to improve performance. This is not only because it redistributes compute resources, but also because it reduces interference between tasks; CPU context switching can have a high overhead cost, particularly due to cache pollution, and task segregation can help reduce that cost. This problem only worsens when you consider the myriad of additional ASIC resources over which there might be contention~\cite{FairNIC2020}.

Secondly, a DPU can be used as an extra security layer. Because it functions as a separate device which can process network packets before forwarding them to the host, it provides an added layer of hardware isolation between the network and the host server, effectively isolating infrastructure and other functions performed on the SmartNIC from those performed on the host, which usually amounts to separating `control plane' operations from `data plane' operations, or infrastructure operations from tenant operations, or network access from user applications, thus further protecting the data center from nefarious tenants. And, with more security operations offloaded to ASICs, one could reasonably say that the more computation one does in silicon, instead of software, the more secure the system will be~\cite{exploringtheadvantagesofdeployingdpus}.

\subsection{Challenges}
\label{sec:challenges}

DPUs and SmartNICs are not flawless solutions, with plenty of challenges of their own regarding their adoption and usage. Here are a few of the issues one may face when working with these devices~\cite{codelineSmartNIC2021}: 

    \begin{itemize}
    
        \item \textbf{Computational Power:}
        DPUs generally have energy-efficient processors that are not as powerful as modern high-end multi-core server processors from Intel, AMD, and NVIDIA; often, the cores of choice are Arm cores. The number of processing cores is also less than that of mainstream server devices. These design choices are motivated by the need to build energy-efficient processors in DPUs, but they increase the difficulty of applying DPUs to some problem domains, as the CPU can easily become a performance bottleneck~\cite{comprehensive_expert_survey}.
        
        \item \textbf{Memory:}
        DPUs often have limited on-chip memory compared to general-purpose CPUs. For instance, a BlueField-2 DPU has 16 GB of main memory~\cite{bf2datasheet_2023}. This forces developers and companies using them to carefully consider how data is managed and accessed, especially for workloads with large datasets. Since DPUs also have limited memory \textit{bandwidth}, high-throughput workloads can also pose a challenge. Efficient data transfer between the DPU and system memory is crucial for preventing bottlenecks; to this end, some DPUs support direct memory access (DMA) between host and DPU.
        
        \item \textbf{Programming:}
        Developing applications and configuring SmartNICs to take advantage of their capabilities can be challenging, requiring that developers be familiar with the individual programming models and APIs of each specific SmartNIC they're working with~\cite{p4programming2013}. Furthermore, their troubleshooting capabilities are often quite limited~\cite{codelineSmartNIC2021}. While some DPU platforms do provide debugging and emulation tools~\cite{comprehensive_expert_survey}, these devices can still generally be difficult or slow to debug, in part because access to the DPU may be indirect.

        \item \textbf{Network Overhead:}
        Off-path devices, specifically, incur a higher network overhead, even if they have a higher overall capacity~\cite{ComprehensiveStudyDPU2023}.

        \item \textbf{Management:}
        Cluster management and debugging is complex, and performance bottleneck debugging with such heterogeneous devices as DPUs can be challenging. From our observations it appears that there may be research gaps around producing or porting better tools for this and related areas.

        \item \textbf{Power and Cooling:}
        Energy efficient as they might be~\cite{SmarterNIC2022, BlockNIC2023, zawawi2023resource}, it is still difficult to add any kind of power-hungry device to pre-existing servers in a data center with power and cooling systems that were not designed for the extra load per-rack, requiring additional considerations for any would-be mass-adopters~\cite{DataCenterDynamics}. As an example, Karamati et al.~\cite{SmarterNIC2022} estimated as much as 6-13\% more power was required for their application, though in practice this can be difficult to measure.

        \item \textbf{Expertise:}
        DPUs are complex to use, and will often require considerable device-specific hardware, IT, and administration expertise.

        \item \textbf{Absence of Standardization:}
        As with any new technology, until regular standards are developed (in the realms of both hardware and software), the variety of SmartNICs and DPUs is broadly heterogeneous~\cite{codelineSmartNIC2021}, meaning that their usage represents a large learning curve with little-to-no flexibility between brands.

        \item \textbf{Financial Cost:}
        SmartNICs and DPUs, being more complex than regular NICs, are more costly in terms of hardware, and also in terms of power consumption.
        
    \end{itemize}

While the research in these areas is perhaps not yet fully mature, there are a number of works which at least attempt to characterize these devices and their idiosyncrasies via benchmarking, as discussed in Section~\ref{section_studies_and_benchmarks}.


\subsection{DPUs on the Market}
\label{sec:dpumarket}
There are currently a variety of SmartNICs on the market which can be classified as DPUs; we review a few important players briefly here, and a few more in Table~\ref{table:dpu-vendors-processors}. Different DPU platforms widely varied and nuanced hardware and accelerator capacities; for deeper critical analyses of how DPU hardware differs we refer the reader to the work by Kfoury et al.~\cite{comprehensive_expert_survey}, and Kashyap et al. and Hu et al.~\cite{kashyap_idiosyncrasies, hu2025dpbento} for detailed idiosyncrasies of BlueField and other DPUs.


\input{Table_2}

\textbf{NVIDIA BlueField}: These DPUs are powerful and provide an optimized infrastructure designed to accommodate a wide variety of workloads in environments such as cloud computing and data centers, achieved through the offloading, accelerating, and segregating of a vast array of workloads. The BlueField-3 is pioneering line-rate packet processing at 400~Gb/s, being the first to boast this capability~\cite{topdpu}. NVIDIA's converged accelerator cards expand upon the BlueField line by adding a GPU, promising heavier in-line computing capabilities~\cite{converged_accelerators}.

\textbf{Marvell OCTEON:} These DPUs are designed with both 5G wireless infrastructure and networking in mind, and are intended to optimize not only cloud workloads, but also machine learning inference on the edge, and other applications. They are developing their own in-house ML engine to work with these devices~\cite{topdpu}. Such hardware-based (ASIC) ML/AI accelerators have a large potential to accelerate AI algorithms, by as much as 100x over software-based implementations~\cite{topdpu, liquid3}, as discussed in Section~\ref{section_ai_ml}. LiquidIO DPUs, like others, are in use by the research community~\cite{xenic2021, comprehensive_expert_survey}.

\textbf{Intel IPU:} The IPU is part of Intel's vision for end-to-end programmable networks, and their target customers are both cloud and communication service providers and enterprises. IPUs incorporate a mixture of ASICs, Xeon CPUs, and FPGAs, and generally have goals similar to other such devices, such as reducing infrastructure overhead and improving security and isolation for both data centers and tenants~\cite{Intelipu, IPU-basedComputingPlatform}.

\textbf{AMD Alveo:} These devices, developed by AMD, are based on FPGAs, making it possible to use them to hardware-acceler\\-ate even custom, computationally demanding tasks such as machine learning inference, data analysis, video transcoding, and more, boasting up to 90x performance compared to a CPU. Given the swifter evolution of algorithms than chip design cycles, the hope is that devices such as this will be able to keep up with algorithmic changes flexibly, which they claim to pioneer~\cite{topdpu}.

\textbf{MangoBoost:} We note this more recent product line as an interesting example of highly differentiated SmartNIC-like devices designed to work together, each accelerating a different aspect of a data center, such as distributed storage, GPUs, or other operations, by offloading network management tasks to them~\cite{mangoboost}. Mangoboost collaborates with AMD, and makes use of Alveo U45N technologies within their own products.


We also note, briefly, the importance of the AWS `Nitro' system and Azure's `Catapult' within the evolution and space of DPU technologies \cite{topdpu, AWSNitro, nitro2023, awsnitrowhitepaper}, but as they are somewhat more proprietary systems, we will consider them less relevant to the current research around using and improving DPUs.

A brief list of a few company whitepapers regarding DPU devices, which the reader may find helpful, can be found in Table \ref{table:benchmarks} of Section~\ref{section_studies_and_benchmarks}; these are included there because they frequently include manufacturer benchmarks.


\section{Programming Languages and Frameworks}
\label{frameworks}

The selection of programming languages, frameworks, and development tools used for SmartNICs varies based on the distinct requirements and functionalities of the SmartNIC, and the requirements of the intended application. The decision regarding the programming languages and tools one will use is often directly influenced by the specific SmartNIC vendor and model; different SmartNICs may possess distinct capacities, and vendors often furnish custom-made, unique software development kits and tools that are meticulously tailored to their hardware platforms---and not others. Furthermore, the programmability attributes of SmartNICs, such as P4 support, are undergoing continual evolution, which also impacts the selection. 

Table \ref{table:toolsets} shows a categorization of various publications based upon the software tools that they use, evaluate, or expand upon in relation to SmartNICs. 

\input{Table_3}

Although other works, such as that by Kfoury et al.~\cite{comprehensive_expert_survey}, go more in-depth, we at least list a number of programming models and toolsets available for programming DPUs, and we also discuss MPI, OpenMP, and gRPC, which they do not, but which are of particular significance regarding parallel and heterogeneous computing:

\begin{itemize}
 
    \item \textbf{DOCA, and other platform-specific SDKs:}
    DOCA is NVIDIA's extensive software development kit (SDK), and more, for BlueField DPUs, and contains virtually all of the important tools and software packages needed for developing applications to work with their hardware, both on the device and on its host, aiming to facilitate and streamline setup and development. It is the go-to for working with NVIDIA DPUs, and is even intended to improve working with AI and ML on these devices~\cite{nvidiaDemystifyingNVIDIA, nvidiadev2023}.

    Other DPU brands often offer similar, extensive platform-specific toolsets, such as AMD Pensando's SSDK~\cite{amdpensando2023}, which are not discussed in this work.
    
    \item \textbf{DPDK:}
    Data Plane Development Kit (DPDK)  can be used in conjunction with DPUs to optimize the performance of high-speed network data packet processing, by offloading network-related tasks from the CPU to DPUs or SmartNICs. This more efficient and rapid handling of network data packets helps achieve both lower latency and higher throughput. DPDK is used in applications such as edge computing, 5G networks, and data centers where speed and efficiency are paramount.
    
    \item \textbf{SDPK:}
    The Storage Performance Development Kit (SPDK) is an open-source framework primarily designed for high-performance storage applications in general, not necessarily for DPUs or SmartNICs. It focuses on optimizing storage operations, especially for SSD and NVMe devices.
    
    However, there is some potential overlap with DPUs. Since DPUs or SmartNICs are hardware accelerators that can handle various network and data processing tasks, in certain scenarios SPDK could be employed with them to enhance the performance of storage-related operations in a network context, particularly when dealing with high-speed data storage and retrieval. The integration and customization of SPDK for use with DPUs or SmartNICs would depend on specific application requirements~\cite{spdkeventframework_2023}.
    
    \item \textbf{MPI:}
    Libraries implementing the Message Passing Interface (MPI) provide general-purpose parallelism without shared memory, MPI itself being a standard which defines interfaces for inter-process communication. Communication in MPI is basically of two types: 1) peer-to-peer communication (sending and receiving messages), and 2) collective communication (e.g., broadcasting a message, or all-to-all patterns). These collective patterns are often necessary when accelerating scientific applications in High-Performance Computing (HPC) clusters. Combining the parallel processing of MPI with DPU offloading, researchers can enhance data processing, simulations, and scientific research in general in HPC, offering the potential for breakthroughs in fields like molecular dynamics, genomics, climate modeling, astrophysics, etc~\cite{bluesMPI2021}. We will discuss works using or expanding MPI in Section~\ref{section_parallel}.
    
    \item \textbf{OpenMP and related models:}
    OpenMP is another programming model widely used in parallel computing, but unlike MPI, OpenMP uses a shared memory paradigm. It allows for the decomposition of sequential programs into parallel components, using compiler directives~\cite{OpenMP2023}.

    Other models and libraries extend this concept of shared memory to work over a distributed system, such as PGAS models, OpenSHMEM, and OpenSNAPI~\cite{gRPC2021}, and works regarding OpenMP will be discussed in Section~\ref{section_parallel}.
    
    \item \textbf{gRPC:}
    Google has developed an open-source Remote Procedure Call (RPC) framework known as gRPC, which boasts high performance capabilities across widely distributed systems. gRPC's remote procedure calls work over standard network connections with a binary protocol, and are language-agnostic. This enables applications to establish seamless communication across diverse systems and languages through a clearly defined interface. While gRPC itself is not specifically tailored for DPUs, it can be deployed alongside them to enhance particular aspects of network communication and data handling. Essentially, it becomes another communication method by which one can offload responsibilities to the SmartNIC, thus decreasing CPU utilization. It is suitable for applications needing rapid data processing and low-latency communication, particularly in data centers and networking infrastructure. Nevertheless, using DPUs and gRPC together will require customized development and configuration in order to make full use of their capabilities~\cite{gRPC2021}. 
    
    \item \textbf{P4}: 
    P4 is a low-level programming language tailored for packet processors, enabling programmers to modify the packet processing routines of switches post-deployment, rendering them adaptable in real-world scenarios. P4 also enables switches to be independent of any specific network protocols. The language has been put forward as a strawman proposal for the future evolution of OpenFlow, a widely used SDN control protocol. P4 has been implemented and evaluated in various scenarios, including dynamic advanced Active Queue Management (AQM) schemes and customized flow tables. It has shown promising results in terms of feasibility, performance improvement, and flexibility for network customization \cite{p4smartnic2020, p4programming2013}.
    
    Expanding P4 has also been explored. Xing et al. \cite{Xing2023} have created a new optimization engine for the P4 compiler which prepares its programs to dynamically rearrange network pipelines during runtime. Their experiments evidenced various successful optimizations in throughput or latency, with improvements ranging from small percentages up to 49\% in one case, with only a small overhead for managing the live optimization software. Their experiments were performed on BlueField-2 and Agilio CX DPUs, as well as a software SmartNIC emulator.
    
    \item \textbf{Hardware Description Languages:}
    FPGAs are typically programmed very differently from CPUs and ASICs, requiring various Hardware Description Languages (HDLs), such as Verilog, VHDL, etc.~\cite{intelIPUprogramming}. This makes the more exotic, FPGA-centric SmartNICs likely more difficult to program than general-purpose CPU-based DPUs, although some systems provide abstractions to remove this barrier, such as Alveo Vitis HLS~\cite{AMDvitisHLS} or OpenNIC shell.

\end{itemize} 


\subsection{Creating New Tools}
\label{sec:new_tools}

A few works create new libraries or APIs. Since a good number of works produce code bases for very specific things, we will primarily mention a few, more general communication libraries here.

Some time ago, Hoefler et al. introduced `sPIN'~\cite{sPIN:High-performance}, a portable packet-processing acceleration programming model that offloads packet processing functions to network cards and enables the acceleration of application and system services, including for redundant in-memory file systems. This acceleration is analogous to compute acceleration using CUDA or OpenCL. They followed more recently by introducing `PsPIN'~\cite{PsPIN2020}, an open-source implementation of the sPIN model for doing computations in the NIC. Their experiments with PsPIN boast 400 Gbit/s packet-processing speeds and low power consumption.

Suresh et al. \cite{NovelFramework2023ieee} present additional API primitives for point-to-point and group communication patterns, for offloading any type of communication pattern to a DPU, in an attempt to combat degraded latency, bottle-necking, and limitations on communication patterns present in other state-of-the-art approaches, such as MPI and OpenSHMEM. They test their work with all-to-all micro-benchmarks, as well as a P3DFFT application.

Other libraries of note, which are detailed in later sections, include AlNiCo, IO-TCP, PEDAL, INEC, DORM, FairNIC, iPipe, and Runway, among others~\cite{Li2022, Taehyun2023, AcceleratingLossy01, inec2020, huang_DORM, FairNIC2020, smartNICsiPipe2019, Runway2023}. 


\section{Parallel and Distributed Computing on DPUs}
\label{section_parallel}

Broadly speaking, parallelism in computing can take a number of different forms, each applicable in often very different situations. We will discuss several relevant, distinct concepts here, and how they apply relative to heterogeneous, parallel, or distributed computing with DPUs. 

\input{Table_4}

Current NIC software development frameworks are typically optimized for ASIC NICs, and do not take into consideration the type of flexibility inherent in off-path SmartNICs' ability to dynamically offload a broader variety of tasks than those defined by a few ASIC accelerators \cite{Xing2023}. Similarly, many parallel computing libraries are not optimized for effective use with these powerful network devices. Some researchers are addressing these problems by adding onto the existing languages and frameworks, creating modifications or compilers specifically for DPU applications, or are re-implementing standards-based frameworks, such as MPI, with SmartNICs in mind. The works outlined below, as well as those following in the remainder of the survey, appear to indicate room for significant software-level optimizations and library-level improvements for SmartNICs. Therefore, in this section we will cite both works expanding upon MPI and OpenMP, and a few prime research examples simply applying various forms of parallelism.

Table~\ref{table:parallel} is intended to provide a useful analysis and comparison of these works' methodologies and contributions. We posit that this table's categorization could provide a framework for the research community to use to develop a more complete taxonomy of research works in this field. Although its scope is too narrow to make broad observations from (having only 12 papers), it can give some quick intuition about each included work's quality or extent of research, and provides a means of finding papers with certain traits.


\textbf{Shared Memory (Thread-Level) Parallelism:} 
This is often done with the aid of libraries such as OpenMP. Because memory is shared, such parallelism is typically only across CPU cores within a single device. As most DPUs are multi-core, a large portion of the works reviewed exhibit this type of parallelism simply in their use of DPU onboard resources.

However, remote direct memory access (RDMA) and network resource allocation systems may be able to circumvent this limitation, allowing for distributed OpenMP systems to work across DPU-networked clusters, as many DPUs support RDMA or similar abstractions.
Usman et al. \cite{OpenMP2023} introduced support for using OpenMP to treat DPUs as network co-processors for offloading, presenting a design in LLVM that supports OpenMP's standard offloading semantics. Their proposed approach offers productivity advantages for programmers working with DPUs, compared to standard OpenMP.

The performance results of Karamati et al.~\cite{SmarterNIC2022} and Kaymak et al.~\cite{derda2023SCposter} demonstrate that it requires considerable engineering effort to achieve worthwhile speedups when using a DPU for applications which are not embarrassingly parallel.

\textbf{Distributed Memory (Device or Process-Level) Parallelism:} Upon adding the limitation that each worker process has its own dedicated, unshared memory, 
work is typically performed on different regions of the data concurrently. Libraries such as MPI allow us to more easily divide and coordinate this work between processes and devices, even heterogeneous devices such as between hosts and DPUs, to perform cluster computing. A number of works combine MPI and DPUs \cite{Large-Message_MPI, bluesMPI2021, SmarterNIC2022, RaDD2020, MolecularDynamics2022, DNN2022}, though any cluster computing done treating DPUs as compute nodes would also fit here.

Most interesting perhaps are the works which not only use MPI, but rather extend or optimize it for use with DPUs, especially to improve upon the collective communication functions, because these require more buffer management and synchronization work among processes, and DPUs are communication-oriented devices that could help address this frequent bottleneck.
In one such example, Sarkauskas et al. \cite{Large-Message_MPI} redesign MPI's non-blocking broadcast and all-gather collectives to make use of DPU offloading, since these primitives are difficult to synchronize efficiently. They test their designs with microbenchmarks on BlueField DPUs, reducing the execution time of \texttt{osu\_ibcast} and \texttt{osu\_iallgather} by up to 54\% and 43\%, respectively.
Additionally, Mohammadreza et al. \cite{bluesMPI2021} proposed a framework named `BluesMPI' that demonstrates a significant enhancement in the overall execution time of the OSU Micro Benchmark when utilising the \texttt{MPI\_Ialltoall} function, with improvements of up to 44\%. The use of BluesMPI results in a notable enhancement in the overall execution duration of the P3DFFT application, with improvements of up to 30\%.
Turalija et al. \cite{MolecularDynamics2022} also intend to extend MPI more fully into the DPU realm with a library they are developing.
%

\textbf{SIMD (Single Instruction Multiple Data) Parallelism:} Expanding upon the prior concepts, SIMD represents a hardware-accelerated application of data-level parallelism, with many small cores performing the same computations, locked in tandem, on different data items. This requires specialized hardware, such as a GPU, TPU (tensor processing unit), vector unit of a CPU, or similar device.

While some DPUs contain SIMD-capable hardware~\cite{converged_accelerators}, and BlueField DPU processors contain SIMD vector units, most DPUs are not targeting GPU-like use cases, instead preferring to be a means of speeding up communication between a datacenter and its host's other hardware~\cite{Lynx2020, HeterogeneousSmartNIC01}, which may include GPUs.

\textbf{Pipeline Parallelism:} Pipelining is the concept of taking a set of instructions that has multiple steps, and instead of completing all steps before moving onto the next data item, stacking them so that any non-dependent steps are performed in parallel, often requiring specialized hardware.

However, pipelining can also be taken advantage of algorithmically across devices, such as when Karamati et al.~\cite{SmarterNIC2022} take an algorithm which is not easily parallelizable under distributed memory settings and rearrange it in order to perform communication-heavy steps on DPUs concurrently with more computational steps on host devices, achieving up to 20\% speedup. Another example is when Kaymak et al.~\cite{derda2023SCposter} perform a complex query's `filter' step on a DPU, and `refine' on its host. One final example is that of Zhong et al.~\cite{LightningZhong}, who also modify an algorithm to effectively pipeline the sending of computations to an optical computing device.

\textbf{Research Gaps:} Some of these works identify areas where further research could be beneficial to the community, beyond what was performed. Some topics of note include HPC offloading using DPUs as low-power accelerators~\cite{OpenMP2023}, more sophisticated halo exchanges~\cite{OpenMP2023, SmarterNIC2022} and using ASICs for communication-intensive simulations~\cite{MolecularDynamics2022}, algorithmic restructurings~\cite{SmarterNIC2022}, in-network runtime software~\cite{RaDD2020}, and the rebalancing of DPU hardware designs~\cite{SmarterNIC2022}.


\section{Broad Studies and Benchmarking of DPUs}
\label{section_studies_and_benchmarks}

In this section, we identify those works which are broad studies, or which perform deeper benchmarks of DPUs. Benchmarking these devices is essential for assessing their performance and efficiency in different applications. Such benchmarks may measure factors such as CPU or network throughput, latency, and energy efficiency, providing insights into the DPU's capabilities under various workloads. These benchmarks help in optimizing hardware and software configurations for specific use cases, ensuring that they deliver optimal performance in their target applications. While a few of the following works are broad studies, and others are in-depth general device benchmarks, other works only measure their specific contribution's effect; since most works do this to some extent, we will identify only a few contribution-specific works of note here. Table~\ref{table:benchmarks} includes a mixture of selections of both.

Note that, due to the relatively small number of works performing thorough benchmarks, we have not specified in Table~\ref{table:benchmarks} which SmartNIC type each work is using, but instead have categorized them by benchmark topic. In this work we do not compile a comparison of benchmarks between available DPU devices, in part because the existing works tend to primarily provide information on only a few more popular models, such as NVIDIA BlueField DPUs; nor do we directly compare those works' results, seeing that their methodology, specific tests or datasets, device or code configurations, and even the ASICS and other aspects of the devices being benchmarked differ so much as to make them incomparable in many cases. Instead, we refer the reader to the well-performed benchmarks described by the papers directly following.

\input{Table_5}

\textbf{Broad Studies and Advice:}
Sun et al.~\cite{ComprehensiveStudyDPU2023} use deeper evaluations of compute power and communication overhead, performed on a BlueField, to recommend four experiment-supported pieces of advice for using off-path SmartNICs to improve performance:
\begin{enumerate}
    \item Use the built-in ASIC accelerators for their intended offloads
    \item In choosing other things to offload, focus on high-latency tasks
    \item Treat the SmartNIC as a separate, independent endpoint, but one with additional resources for the server host to use
    \item Do not simply use the same system design as you would for an on-path SmartNIC, as it often won't perform as well as an on-path device in those areas.
\end{enumerate}
Finally, they discuss possible refutations of performance gains shown by a few other papers, demonstrating the difficulty of creating truly effective offloads for off-path SmartNICs.
%
Wei et al. \cite{CharacterizingSmartNIC2023} also perform a comprehensive study of the BlueField-2, focusing on file system and key-value store (KV-store) communication paths, and using multiple paths concurrently, to discover notable performance improvements. They then give advice on finding such potential optimizations, which they have validated through their studies. 
%
J. Liu et al. \cite{PerformanceCharacteristics2021} also evaluate the capabilities of the network and computing aspects of the NVIDIA BlueField-2, and showed that while the device is flexible, it is easy to bottleneck and overwhelm it. They recommend a selection of operations for offloading, which this DPU performs particularly well at. Kashyap et al.~\cite{kashyap_idiosyncrasies} focus on evaluating the DMA and RDMA paths of multi-generational DPUs, and usefully identify a number of device idiosyncrasies.

Chen et al. \cite{chen2024datapathaccelerator}, instead of measuring the capabilities of the CPU, delve into the datapath accelerator (DPA) of the BlueField-3, which is a small real-time unit with low resources but a high thread count, primarily intended to accelerate the DPU's NIC system. They find it can be efficiently applied to certain types of offloads, and give advice thereon.

Finally, recent detailed surveys can be very instructive, such as those presented by Kfoury et al.~\cite{comprehensive_expert_survey} for general information about SmartNICs, and Nickel et al.~\cite{in_network_computing_survey} for information on in-network computing with SmartNICs. Works discussing benchmark suites may also give broad useful information, such as that given by `dpBento'~\cite{hu2025dpbento} regarding the types of tasks DPUs perform most effectively.

\textbf{A Big-Picture View on Cloud Architecture:}
Some years back, Caulfield et al. \cite{Caulfield2018} published an invited work which is not so much a broad study or benchmark, as it is an analysis of the current state of the research regarding fully programmable networks. They provide an in-depth discussion towards applying FPGA SmartNICs to create a more flexible cloud infrastructure, going beyond the use of SmartNICs for offloading specific tasks, and discussing the architecting of highly dynamic data center systems to meet the growing needs for abstractions able to power infrastructure, applications, and developers alike. We find this discussion interesting, as it appears to reflect how network hardware and software have developed since its publication; for example, they lack reference to proper DPUs, which now possess some of the abstractions the paper calls for. They provide an informative big-picture view.

In a more recent review, Grant et al. \cite{RaDD2020} discuss the design of SmartNICs and how they can be used to offload current runtime software, as well as enable future in-network runtime systems.
D{\"o}ring et al.~\cite{smartnics2021} discuss the demand for these devices, as well as their creation and use in attempting to meet the growing needs surrounding Software-Defined Networking (SDN). They highlight how SmartNICs can be used to tackle the lack of network flexibility, making virtualized networks more feasible. The paper also aims to provide a universal definition for the term ``SmartNIC''. 
%
Barsellotti et al. \cite{dpuedge2022} go on to discuss three different use cases for DPUs at the edge: network monitoring for 5G networks, power efficiency in edge-to-cloud, and added network security.
%
Horany et al.~\cite{gRPC2021} recommended and describe a model for offloading and allocating SmartNIC resources across a network, using gRPC, among other tools.

On a related note, Justine Sherry \cite{sherry2024driven} makes an argument that instead of considering DPUs only for compute acceleration and offload, these devices can develop to become the overall data movement controller (DMC) for a compute node with a multi-core CPU and other accelerators like GPUs. This is a new paradigm for the role of emerging DPUs, compared to traditionally making the host responsible for both computations and orchestrating data movement. The paper presents challenges and open questions in realizing this vision in the data center environment, including a list of unresolved queries regarding how the hardware and software systems of the future will adapt to accommodate such a dedicated `NIC-DMC' operating independently from the CPU complex.

\textbf{Compression Studies:}
Recently, there have been a number of works studying, more deeply, the benefits of using DPUs in order to accelerate data compression tasks, utilizing the compression related hardware building blocks in NVIDIA's BlueField SmartNICs. One particularly relevant work is that of Li et al.~\cite{LossyandLosslessCompression02}, who first examine the performance of lossy and lossless compression on NVIDIA BlueField DPUs (particularly SZ3, DEFLATE, and zlib) using seven real-world datasets, underlining the potential of offloading compression tasks from host CPUs to DPUs to enhance the performance in data-intensive applications. Later, they move to further accelerate such compression tasks, which we will describe in Section~\ref{section_comm_infra}, along with a number of other works on compression.

\textbf{Creating New Benchmark Suites:}
And finally, where there are works using benchmarks, there are also works creating new benchmarks.
`dpBento'~\cite{hu2025dpbento} is a benchmark suite specializing in data processing; its creators share helpful insights found by their own use of it. We find this paper to be of particular interest because of these insights; for example, they observe that some DPUs may outperform their hosts at processing smaller operands and floating-point operations, which may be of interest to those offloading some types of data-oriented tasks, such as the geospatial queries of Kaymak et al.~\cite{derda2023SCposter}, discussed later.
Wang et al. \cite{wang2023dpubench} proposed `DPUBench' an application-driven benchmark suite, covering three common categories of DPU applications: network, storage, and security. The suite includes a scalable framework for benchmarking how well DPUs handle tasks relating to those categories.
Nearly of the same name, Michalowicz et al.~\cite{dpubench2023} created a micro-benchmark suite, also called `DPU-Bench', for helping HPC researchers to determine how many processes can be offloaded to a DPU without degrading performance, given various factors.
Kashyap et al.~\cite{kashyap_idiosyncrasies} create a microbenchmark, DPUDMABench, for benchmarking DMA capabilities.


\section{Application Areas of DPUs}
\label{apparea}

The remainder of this survey might be considered a form of topical index, intending to point researchers to more complete treatments of each area than we can give here, categorizing and grouping the main body of papers we have reviewed by the various applications to which DPUs and SmartNICs are being applied. Additionally, we have separated out those works connecting these devices with HPC, as well as various forms of machine learning (ML) and artificial intelligence (AI) applications, as separate areas of interest.

Since the line is often very blurred between infrastructure and more unconventional applications of SmartNICs, in this section we will also present those works which experiment with using SmartNICs for various other kinds of offloads and co-processing tasks, reaching far into the realm of applications as well. Although they are quickly becoming crucial in infrastructure acceleration, the flexibility of DPUs has led to their skyrocketing adoption in a surprising range of domains beyond that niche. From blockchain technologies to molecular dynamics simulations, they are being treated as co-processors, data management and analytics devices, AI-accelerators, and more. Table~\ref{table:infr_and_app} gives a reasonable breakdown of the many overlapping topics, and their sources, within this category.

\input{Table_6}


\subsection{DPUs in Commercial and Infrastructure Applications }
\label{section_comm_infra}

The commercial application of DPUs has become increasingly prevalent in modern computing environments. As they are primarily infrastructure devices, much of the current research around DPUs is in using them to improve the infrastructure itself of the data centers they belong to; their ability to accelerate network communication and offload tasks relating to packet processing, security, infrastructure processing, storage, virtualization, cloud computing, etc. can provide an important advantage to the companies that use them. Since these are the primary familiar and intended applications of SmartNICs, these areas of research tend to show many positive results, as described in the papers we address in this section.

\textbf{Network Offloading and Load-Balancing Frameworks:}
`DORM' is a recent framework by Huang et al.~\cite{huang_DORM} which performs optimization on load-balancing.
M. Liu et al. \cite{smartNICsiPipe2019} presented `iPipe', an ``actor-based'' framework to aid in offloading varied, distributed applications to SmartNICs in a way that is load-balanced, despite differences in task execution costs. Evaluations of iPipe show core savings (up to 3.1 decent Intel cores in a 10Gbps scenario) and latency reductions by 23.0 $\mu$s when offloading to SmartNICs for data analytics, transaction systems, and KV-stores. However, \cite{ComprehensiveStudyDPU2023} questions the efficacy of iPipe’s design for off-path SmartNICs.
%
Ravi et al. \cite{Runway2023} introduced the `Runway' framework, which uses an object storage abstraction to allow for computation on data while it is in transit, and offloading compression to a heterogeneous device. This framework is specifically designed to be adaptable in order to execute user-defined functions during runtime. They perform tests on both CPUs, DPUs, and GPUs.
%
`DPDPU'~\cite{CIDR25} focuses on gracefully handling the heterogeneity of DPU-networked systems' networking, compute, and storage in data processing tasks. The framework `PsPIN', previously mentioned in~\ref{sec:new_tools}, is also worth noting here. Finally, Fuhrer et al. \cite{latency01} use on-NIC reinforcement learning for network congestion control, but we will discuss this work in Section~\ref{section_ai_ml}.

A potential way to use SmartNICs to improve disk and network I/O operations is by rearchitecting the TCP stack itself, as done in \cite{Taehyun2023} by Kim et al. They present `IO-TCP', which improves content delivery by splitting its stack of operations between the host CPU and the SmartNIC. Since offloading the whole stack to the SmartNIC would leave data inaccessible to the CPU, and the SmartNIC is not fast enough to compete for some operations, they differentiate between ``control plane'' operations to be handled by the CPU, and ``data plane'' operations to be managed by the SmartNIC. This reduces cache pollution between operations with large amounts of disk I/O. Their work is tested on the BlueField-2 DPU, demonstrating approximately double the throughput in cases with high CPU congestion, even compared to systems using far more CPU cores.

We note that many of the works below also bear domain-specific improvements in networking and communication efficacy, but we have chosen to categorize them by their particular application domain where sensible, rather than by their application to the very broad category of ``networking''.

\textbf{Hardware and FPGA-Based; Network, Packets, \& Transactions:}
Some works, such as that of Zhang et al. in \cite{Zhang2023}, aim to improve the design of SmartNICs themselves. Their efforts demonstrate `SmartDS', an FPGA prototype SmartNIC with the ability to split message headers and bodies between the host and SmartDS, offloading compression and improving throughput for distributed storage systems, with potential for accelerating and reducing the number of middle-tier servers required for cloud storage systems, etc. Their tests show up to 4.3x throughput and a latency reduction of 2.6x.
%
Similarly, Li et al. develop `AlNiCo' \cite{Li2022}, an on-NIC network transaction scheduling system which aims to offload the scheduling of transaction requests and responses to specific host CPUs, in a manner which reduces both overhead and resource contention, and which adapts to feedback during dynamic workloads. Their solution uses an FPGA-based SmartNIC to program their own hardware acceleration, into which they feed feature vectors comprised of request information. They also use RDMA (Remote Direct Memory Access), and the development of their own ``scheduling-enabled'' RPC (Remote Procedure Call). They show improvements to both throughput and latency.
%
Also, Yizhou et al. \cite{SuperNIC} fill a niche by presenting what they claim is the first SmartNIC that is both multi-tenant for virtualization needs, and programmable but hardware-based, which they call `SuperNIC' (in a naming collision with NVIDIA's SuperNIC products). They focused on optimizing execution of entire connected groups (flows, or directed acyclic graphs) of network tasks, and included various optimization techniques resulting in low overhead and high throughput compared to the baseline. Furthermore, the researchers proposed a fair sharing mechanism for the SuperNIC's hardware resources.
%
As for traffic analysis, Siracusano et al. \cite{Siracusano2022} use a hardware-implemented binary neural network, described in section~\ref{section_ai_ml}), to speed things up by two orders of magnitude. 
Finally, Brunella et al.'s `Hyperion'~\cite{HyperionAC2022}, described later, is a more direct example of modifying DPU hardware, in pursuit of better distributed storage systems. 

\textbf{Distributed File Systems:}
Others have also been able to achieve substantial speedups by focusing on distributed file systems (DFS).
LineFS, a persistent memory (PM) distributed file system (DFS),  was first presented by Jongyul Kim et al. in a paper describing its design, implementation, and evaluation~\cite{Kim2021}. LineFS resides within the SmartNIC itself to offload DFS operations and reduce CPU contention, while still taking advantage of new client-local persistent memory (PM) to create speedups by preferring client-local behavior. They test their work on a BlueField and compare it to Assise (a current PM DFS), showing improvement in DFS availability, latency of LevelDB operations by up to 80\%, and throughput in Filebench by as much as 79\%. However, \cite{ComprehensiveStudyDPU2023} notes a limited range of situations in which LineFS improves performance.
%
Wei et al. \cite{CharacterizingSmartNIC2023} perform case studies using both LineFS (an on-SmartNIC DFS) and DrTM-KV (an RDMA-based disaggregated KV-store), focusing on the management of such systems, and the idea of using multiple communication paths concurrently, achieving up to 30\% and 25\% improvements, respectively, using a BlueField-2 DPU.

Gootzen et al. \cite{dpfs2023} propose `DPFS', which represents a decoupling of the host from its file system and using \texttt{virtio-fs} to virtualize it onto the DPU. The host's computation is freed up significantly, with little latency cost for tenants. 
K. Zhong et al. likewise proposed `DPC' \cite{DPC01}, a high-performance, DPU-accelerated, distributed/standalone file system client, designed to offload computationally intensive tasks to DPUs. DPC optimizes file operations through `nvme-fs', an enhanced NVMe protocol that enables low-latency and high-performance interactions. Additionally, they introduce `KVFS', a key-value-based file system that replaces underutilized local disks with disaggregated storage. Experimental results show significant improvements, including over 80\% CPU usage savings in high-concurrency scenarios, and a 90\% CPU usage reduction in distributed file systems. 

The conceptual design itself of a DPU is altered by Brunella et al.'s `Hyperion', which replaces host CPU-based devices as the primary manager of distributed NVMe storage~\cite{HyperionAC2022}, reimagining typical architectures to eliminate the need for many CPU server devices.

\textbf{Compression:}
As some DPUs contain compression ASICs, it is demonstrable that compression time and latency can be drastically reduced via their use.
Li et al. \cite{AcceleratingLossy01} introduced a library called `PEDAL' that optimizes data compression designs utilizing the DPU’s hardware capabilities. They also highlight the challenges of existing compression methods due to high computational demands, and present a co-design with the MPICH MPI library which achieves a speedup of up to 101x in compression time, also reducing the communication latency by up to 88x, compared to not using ASICs. Lastly, they focused on enhancing the data compression in communication-oriented HPC scenarios. 
Ding et al. \cite{D2Comp2024, DComp2023} also choose to offload compression to the DPU with `D$^2$Comp', in order to accelerate LSM-tree compaction; Ravi et al. \cite{Runway2023} do so with `Runway' for in-transit computation; and Zhang et al. \cite{Zhang2023} with SmartDS (all further detailed in other parts of this survey). Others, such as Jianshen~\cite{ProceedingswithsmartNIC}, make use of the compression ASICs as well.

\textbf{KV-Stores, Databases, \& Big Data Systems:}
SmartNICs are being used to offload various aspects of distributed data stores, with many claims that these offloads appear to be effective.
In \cite{Lin2023}, Lin et al. focus their efforts on the shuffle processes (map-reduce, etc.) of data-intensive applications, and more generally on dynamically partially offloading operations requiring intermediate data exchange between compute nodes. They present `SmartShuffle' for coordinating offload to maximize usage of both the CPU and SmartNIC simultaneously. They attach it to Spark and test it on a Broadcom Stingray DPU, achieving up to 40\% faster benchmark performance than Spark.
%
Somewhat related, Schuh et al. \cite{xenic2021} introduce `Xenic', a system that employs an asynchronous, aggregated execution paradigm with adaptable point-to-point communication between SmartNICs, improving both network and core efficiency in performing sharded data store transactions. Their experimental findings demonstrate that Xenic achieves notable improvements on three different benchmarks, at least doubling throughput and reducing latency. However, \cite{ComprehensiveStudyDPU2023} questions Xenic’s use in off-path SmartNICs.

As for more specific aspects of such systems, Ding et al. \cite{D2Comp2024, DComp2023} proposed another improvement, `DComp' and then `D$^2$Comp', wherein they integrate DPU-offloaded LSM-tree compaction with RocksDB. This approach was evaluated using on a BlueField-2, accelerating compaction performance by up to 4x, write and read throughput somewhat, and reduce host CPU contention.
%
Also, Shi et al. \cite{inec2020} proposed a set of coherent in-network, tripartite graph-based erasure coding (EC) primitives, named `INEC' (building on their previous work named `TriEC'~\cite{triec2019}), to more easily allow offloading of EC to SmartNICs despite the variety of schemes EC follows. The researchers demonstrated that NIC implementations of INEC primitives, for various EC schemes, with a co-designed KV store, results in improved throughput and write performance.

Lastly, when dealing with network data flows through SmartNICs, Liu et al. \cite{ProceedingswithsmartNIC} recommend the use of Apache Arrow at the foundation, because its data format can save data transformation time. The authors share their experience in adjusting a partitioning algorithm for particle data to work with Apache Arrow. 

Works belonging to other sections, such as that by Wei et al.~\cite{CharacterizingSmartNIC2023}, may also deal with KV-stores.

\textbf{5G Network Infrastructure:}
Yan et al. \cite{p4smartnic2020} introduced a SmartNIC solution that is P4-enabled and FPGA-based. This solution was specifically developed to cater to the networking requirements of web-scale cloud and the 5G / beyond-5G era. The authors demonstrated the application of this solution in a 5G environment, with a particular focus on network slicing from edge data center to core data center. Their experiments achieve a throughput of up to 84.8 Gbps, even using only a single CPU core, and up to 30\% higher bandwidth.
%
Later, Borromeo et al. \cite{FPGA2022} add a 5G DU Low-PHY layer to an FPGA SmartNIC using the OpenCL framework. They demonstrate that incorporating these 5G functionalities into a SmartNIC as an offload, as opposed to other current solutions for a 5G network, can result in notable reductions in both processing time and power consumption.

\textbf{Multiple Tenants, Virtualization, and SDN:}
Applications running on SmartNICs have been limited regarding tenants, because they require a level of mutual performance isolation which is unsupported by current software~\cite{FairNIC2020}; when the available resources include ASIC hardware accelerators in addition to the CPU, this complicates potential sharing mechanisms. `FairNIC', created by Grant et al.~\cite{FairNIC2020}, provides this isolation between tenants, enabling multiple applications to co-habitate with fair resource sharing on a single NIC without impacting each other's performance, and demonstrating that sharing SmartNICs among virtual tenants is feasible, though it will require appropriate security mechanisms. 
%
`DORM' attempts to improve upon such works, using optimization schemes to load-balance between the SmartNIC and host.
We also note Yizhou et al.'s~\cite{SuperNIC} work on `SuperNIC', also towards multi-tenancy and virtualization, which we have described previously.
Partially related, Njavro et al.~\cite{dpusolution2022} apply DPU offloading to container overlay networks to decrease congestion in the host.

In an older work, Yanfang et al. \cite{uno2017} put forward an architecture, named UNO, to provide SDN-controlled network function offloading through a virtual management plane, using multiple switches within the host, without interfering with data-center-wide data-- and control-planes. The experimental findings reduce control-plane overhead, and demonstrate a potential reduction in power consumption by a factor of 2.

\textbf{Security:}
Kiraf et al. \cite{dpunumerousvpnconnection} investigate the potential of DPUs as a scalable and transparent security solution. This solution incorporates Intrusion Detection Systems / Intrusion Prevention Systems (IDS/IPS) and supports multiple VPN connections. The study primarily focuses on two aspects: first, the ability of the DPU to process, filter, and route all data to and from the OS, thus reducing its workload and enhancing security; second, the scalability of the solution. The experimental results demonstrate effective use of the BlueField-1 DPU to ensure a transparent and scalable security solution in terms of VPN connections. 
%
Also, Miano et al. \cite{DDoS2019} direct the reader through the process of using a SmartNIC to offload mitigation of Distributed Denial of Service (DDoS) attacks, taking advantage of on-NIC hardware filtering.

\textbf{Kernel:}
Ji et al. \cite{styx2023} proposed STYX, a framework that offloads memory optimization features from the Linux kernel to SmartNICs. It reduces the disruption of application execution, improving data center efficiency. STYX utilizes the SmartNIC's RDMA capability to copy memory regions, resulting in a 55-89\% decrease in 99th-percentile application latency\cite{styx2023}. 

\textbf{Blockchain:}
While the use of blockchain technology has become widespread across various industries, it is widely acknowledged that blockchains have significant potential environmental impact due to their high energy consumption and hardware requirements~\cite{BlockNIC2023}. In a slight reduction of this impact, Kapoor et al.~\cite{BlockNIC2023}, building on their previous work~\cite{BlockchainSmartNIC2023}, presented a blockchain infrastructure that runs completely on SmartNICs to make use of spare processing power, called `BlockNIC'.



\subsection{DPUs in HPC and Scientific Applications}
\label{section_HPC}

In the realm of high-performance computing (HPC) and scientific simulations, a number of works have begun experimenting with using DPUs both to improve HPC systems, and to offload calculations, such as with large simulations---in particular, for tasks which require heavy communication patterns, such as `halo exchanges', as found in molecular dynamics (MD) and particle simulations.

However, for such offloads to be effective, one often must start thinking outside the box; many algorithms are designed for certain architectures, and so it may often be beneficial to actually restructure current algorithms.

\textbf{Molecular Dynamics Simulations:} In one prime example of algorithmic restructuring, Karamati et al.~\cite{SmarterNIC2022} evaluated the benefits of using NVIDIA BlueField-2 DPUs as compute accelerators for MiniMD molecular dynamics simulations (a proxy for LAMMPS software). They redesign an existing algorithm to perform pipeline parallelism, which highlights the necessity of rearranging some algorithmic patterns to better take advantage of parallelism and DPUs' strengths and communication patterns. They provide valuable advice to this end. With their restructuring, the BlueField-2 provided up to 20\%  speedup, with no real loss of simulation accuracy.
%
Additionally, Turalija et al.~\cite{MolecularDynamics2022} also proposed DPUs for innovative methods of MD simulation acceleration, inspired by the Anton supercomputer's specially designed ASICs, since DPUs also allow access to networking ASICs, which can be used to accelerate MD simulations in similar ways to Anton. This opens the door for other systems to potentially reach the same microsecond timescale needed for many real-world applications.

In a similar vein, Ulmer et al.~\cite{composableDataServices2023} extend data services onto SmartNICs in order to leverage resource isolation, also for the benefit of HPC applications, such as particle simulations. The paper addresses how to construct software for implementing services on SmartNICs, and seeks to learn how useful they will be. Experimental results from a 100-node cluster using BlueField-2s indicate utility in data management tasks, though they are, as expected, less performant than their hosts.
%
Likewise, Liu et al.~\cite{ProceedingswithsmartNIC}, described previously, perform experiments regarding particle data. Both these works use Apache Arrow.
%

\textbf{HPC Systems:} Horany et al.'s work~\cite{gRPC2021} aims to help with allocating network-wide resources for large-scale systems. Tong et al.'s work~\cite{tong_DALdex} on CPU-DPU hybrid learned indexes also targets the needs of HPC systems, and Suresh et al.~\cite{suresh_krylov} lay the groundwork for offloading vector and matrix operations and test it on Krylov solvers.

\textbf{Geospatial Computing:} Kaymak et al.~\cite{derda2023SCposter} experiment with offloading the `filter' step in spatial join operations (polygon intersection queries), leaving the heavier `refine' step computations to the host devices. They use a cluster of DPUs for their experiment, showing some performance improvement. They also briefly measure scalability.


\subsection{DPUs in AI and ML Applications}
\label{section_ai_ml}

Due to the current importance of artificial intelligence (AI) and machine learning (ML) in both research and industry, it is of particular interest that we note how DPUs have been used in conjunction with these technologies. Often, this means taking advantage of FPGAs to create hardware implementations of neural network architectures or operations, which can sometimes boast over two orders of magnitude faster inference \cite{latency01, Siracusano2022, LightningZhong}, although it can also simply be done via computational offloading. See Table~\ref{table:neural_net_type} for a breakdown of the types of neural networks etc. that the research is exploring.

\input{Table_7}

\textbf{Making Deep Learning (DL) Scalable:}
There are various problems with scaling DL systems, which may be addressed via clever use of SmartNICs.
Guo et al. \cite{FCsN_NN_FPGA} proposed a framework called `FCsN' for performing neural network inference on FPGA-based SmartNICs, aiming to improve the performance of HPC and data center processing. Their experimental results, tested on both DNNs and Graph Neural Networks (GNN), boast up to 10x speedups compared with a baseline of using MPI on CPUs.
Later, Guo et al. \cite{HeterogeneousSmartNIC01} continue their work, proposing a heterogeneous SmartNIC system with both hardware and software co-designed, for distributed deep learning recommendation models (DLRMs). Their intent is to resolve the all-to-all communication bottleneck these models suffer from and increase their scalability. Their approach enhances locality and computational efficiency, achieving a 2.1× inference latency speedup, and a 1.6× training throughput speedup. Both these works measure how their contributions affect scalability.

Zawawi~\cite{zawawi2023resource} identifies a ``data stall'' issue, where extensive data preprocessing at the CPU during DL training causes host devices to struggle to keep up with the training GPUs, causing the whole pipeline to stall and GPUs to spend time idle. They help mitigate this by offloading some of this preprocessing to the DPU, also somewhat improving energy efficiency within the data center by nature of the lighter-weight DPU energy requirements.
On a similar note, Tork et al. \cite{Lynx2020} presented `Lynx', a new network server architecture for offloading both data and control planes of network tasks to SmartNICs. The design is accelerator-centric, capable of running without CPU help by enabling more direct networking to the GPUs. This attains greater than 4x throughput for their example GPU task of face-verification, and 25\% increase for an accelerated inference task, compared to host-centric architectures. However, \cite{ComprehensiveStudyDPU2023} questions the efficacy of running Lynx on a SmartNIC versus the host.

\textbf{AI/ML for Network Traffic Management}:
Fuhrer et al. \cite{latency01} introduce a lightweight congestion control solution for datacenters using a reinforcement learning algorithm, RL-CC, which is converted into decision trees to achieve a 500x reduction in inference time. Deployment on NVIDIA ConnectX-6~Dx SmartNICs demonstrated superior performance in managing bandwidth, latency, and packet loss in a balanced way, compared to existing algorithms like DCQCN and Swift, across various benchmarks.
On a related note, Hardware-based, on-NIC Binary Neural Networks (BNNs) are introduced to the data plane of FPGA SmartNICs by Siracusano et al. \cite{Siracusano2022}, in order to do classification and analysis of network traffic at a line rate of 40Gbps using machine learning. This hardware-accelerated neural network drastically improves latency compared to software implementations of similar networks (by up to two orders of magnitude), and appears to achieve decent accuracy, while freeing up the host's CPU for other work. Their work demonstrates significant value in researching hardware implementations of neural networks.

As for software implementations, Tasdemir et al. \cite{inventifationMLBluefieldDPU} investigate deploying Machine Learning (ML) algorithms on BlueField-3 DPUs for SQL injection detection. They test 20 different ML models on an SQL dataset, achieving near-real-time detection with a Passive Aggressive Classifier, with an accuracy of 99.78\%.

\textbf{Co-Compute Offloading:}
Arpan Jain \cite{DNN2022} investigate the potential benefits of using the Arm cores of BlueField-2 DPUs to expedite the training of DNN models by offloading phases of the training to them; their study may represent the first such attempt. MPI is employed despite the heterogeneity of devices, and the researchers' approach ultimately demonstrates up to 17.5\% improvement in training duration on cutting-edge HPC clusters. They try several designs, and test them on multiple DL model types. They also measure scalability.
Likewise, Shibahara et al. \cite{federatedlearning} point out the heavy network processing load for Federated Learning schemes, which require aggregating local weights from distributed learners and then sending global weights back out. They make use of the DPU's cores to offload this aggregation, with a 1.39x speedup.

In another offloading effort, Perea-Trigo et al.~\cite{DPUweapondetectioncase2022} investigate the potential advantages of employing an NVIDIA BlueField-2 DPU to mitigate the burden experienced by a continuously operating server. They quantify the extent to which a DPU can alleviate the workload of a CCTV system for weapon detection, evaluating its performance across various scenarios. They observed a remarkable reduction in workload of 43,123\% over a 24-hour period, with savings exceeding 98\% during nighttime intervals~\cite{DPUweapondetectioncase2022}.

Also, Tootaghaj et al. \cite{EdgeForElasticity} proposed a new architecture, `Spike-Offload', which strategically offloads workload spikes from microservices (important for edge ML workloads), when the SmartNIC has available compute power, thus reducing service level agreement (SLA) violations and giving better performance, lower energy consumption, and 40\% potential reduction in capital expenditure. 


\textbf{Optical-Electronic Technologies:}
Z. Zhong et al. \cite{LightningZhong} proposed `Lightning', a hardware-based prototype SmartNIC with photonic computing cores. Lightning is a pioneer in leveraging the fast data paths of a SmartNIC to overcome the bottleneck of controlling and moving data from the electronic domain to the photonic domain, where DNN inference can be performed much faster. To achieve this, they modify an algorithm to be more pipelined. Their prototype and simulations are capable of doing DNN inference in real-time, outpacing the same inference done purely on NVIDIA A100 GPUs by over 300x (and thus also using 300x less power).







\section{Conclusion}
\label{conclusion}
Over the course of our research for this survey, a couple of points have become clear, so that we are able to draw a few broad conclusions from this body of work.

First, this is a very active field of research, and a large number of works demonstrate achievable performance improvements using SmartNICs, offloading a broad variety of types of work. In areas where there is potential for future compatibility, we recommend that the members of the research community pursue integration of their solutions, being thoughtful of eventual means of combining their methods with others', in order to maximize the potential impact of their work and broaden its scope beyond their particular use-cases. Such collaboration likely has the potential to result in impressively compounding improvements. Even in cases where the same resource is maximized by two or more solutions, such as the many works which achieve performance boosts by filling the DPU processors to capacity, concepts from individual solutions might be combined into a single integrated solution. At the very least, the community would likely benefit from some form of collaborative, well-supported, diverse, cohesive suite of such solutions, each with its varied strengths, to accessibly choose from when developing with DPUs. Even among the SDKs provided by DPU vendors, standardization and cross-compatibility are currently lacking~\cite{comprehensive_expert_survey}.

Second, it is clear from a large segment of the research that the largest speedups and performance improvements shown are almost exclusively those which come from taking advantage of hardware acceleration: making use of the on-board ASICs, specialized FPGA programming, laying the groundwork for photonic computation, etc. These hardware-leaning efforts continue to be the results which most often make orders of magnitude of difference, compared to current methods. We thus recommend searching for every possible opportunity to apply the built-in accelerators to your target task. Standard parallel computational offloading may still be worthwhile, and likely energy-saving, though it is usually capped by the computational capacity of the DPU and the effects of reducing cache pollution and CPU congestion in the host.

Finally, much research is currently being done in exploring DPUs and SmartNICs as heterogeneous co-processors and accelerators, as opposed to simply as data center infrastructure offload and management devices, for nearly any kind of work-offloading or benchmarking. Much of this research appears to be yielding favorable performance improvements, ranging from meager to large. Interest in these heterogeneous computing systems has even spread to diverse fields outside of computer science and IT, such as the natural sciences and engineering. Because such on-NIC computational applications are not the primary target of most SmartNICs, these research angles provide momentum to the field in an interesting new direction; likewise, we see the development of DPUs with higher and higher compute capacity mirroring (or perhaps causing) this trend. New frameworks are being developed to provide the abstractions needed to bring these devices into wider usage, making them increasingly accessible and effective. Despite this progress, several obstacles persist, but researchers appear to be working to address these challenges as the niche for DPUs continues to develop and evolve.

\section{Acknowledgment}  
This material is based upon work supported by the National Science Foundation under Grant No. (2344578). Any opinions, findings, and conclusions or recommendations expressed in this material are those of the author(s) and do not necessarily reflect the views of the National Science Foundation.



\bibliographystyle{elsarticle-num}
\bibliography{citations}






\end{document}

%% file: Table_1.tex
\begin{table}[h]
\centering
\small
\begin{tabular}{lll} \hline 
    \textbf{Characteristic / Offload Ability} & \textbf{On-Path} & \textbf{Off-Path} \\ \hline
    
    Communication Overhead                    & Low              & High              \\
    Operating System                          &                  & \checkmark        \\
    Low-Level Control                         & \checkmark       &                   \\ \hline
    
    ASIC Offloads                             & \checkmark       & \checkmark        \\
    App. Offloads \& Co-Processing            & If CPU/FPGA      & \checkmark        \\
    
    \hline
\end{tabular}
\caption{Characteristics of on-path and off-path SmartNICs.}
\label{table:data_path}
\end{table}

%% file: Table_2.tex
\begin{table}[h]
\resizebox{\columnwidth}{!}{ 
\begin{tabular}{llll} 
\hline
    \textbf{Vendor}   & \textbf{Product Line}   & \textbf{Product Example} & \textbf{Core Type} \\ \hline
      
    NVIDIA Mellanox   & BlueField               & BlueField-2, 3           & CPU                \\
    NVIDIA            & Converged Accel.        & AX800, A100X, ...        & CPU+GPU            \\
    Marvell           & OCTEON, ARMADA          & LiquidIO III, ...        & CPU                \\
    Intel             & IPU                     & Mount Evans, ...         & CPU+FPGA           \\
    AMD (Xilinx)      & Alveo                   & (several)                & FPGA               \\
    AMD (Pensando)    & ---                     & Giglio, Elba, ...        & CPU                \\
    Broadcom          & Stingray                & (several)                & CPU                \\
    Netronome         & Agilio CX               & (several)                & CPU                \\
    Achronix          & Speedster7t FPGA        & (several)                & FPGA               \\
    Fungible          & ---                     & F1, S1                   & CPU                \\
    MangoBoost        & ---                     & (several)                & (various)          \\ \hline
\end{tabular}
}
\caption{A few major players in the DPU market, both vendor, product line, and processor technology. This information was mostly gathered from Asterfusion~\protect\cite{topdpu}, general knowledge, and product datasheets~\protect\cite{converged_accelerators}.}
\label{table:dpu-vendors-processors}
\end{table}

%% file: Table_3.tex
\begin{table}[h]
\centering
\small
\begin{tabular}{lllll} 
\hline
    \textbf{Software} &  \textbf{Related Works}    \\ \hline
    DOCA               & \cite{smartNICsiPipe2019, MolecularDynamics2022, nvidiaArch, dpubasedhardwareacceleration}  \\ 
    DPDK / SPDK        & \cite{styx2023, xenic2021, FairNIC2020, PerformanceCharacteristics2021,  Runway2023, smartnics2021, DDoS2019, federatedlearning, ProceedingswithsmartNIC,                                           wang2023dpubench}   \\ 
    MPI                & \cite{Large-Message_MPI, bluesMPI2021, SmarterNIC2022, RaDD2020, MolecularDynamics2022, DNN2022, NovelFramework2023ieee, nvidiaArch, suresh_krylov}     \\ 
    OpenMP             & \cite{OpenMP2023}    \\ 
    gRPC               & \cite{derda2023SCposter, gRPC2021}        \\ 
    P4                 & \cite{p4smartnic2020, zawawi2023resource, Xing2023}  \\ \hline
\end{tabular}
\caption{Classification of publications based on software used (frameworks, libraries, and languages). These works may make use of the software, evaluate it, or expand upon it, in SmartNIC applications. Many other works may use libraries we were unaware of.}
\label{table:toolsets}
\end{table}

%% file: Table_4.tex
\begin{table*}[h]
\centering
\small
\resizebox{\textwidth}{!}{ 
\begin{tabular}{lccccccccccccc} 
\hline
                                                                 & \cite{OpenMP2023} & \cite{Large-Message_MPI, bluesMPI2021} & \cite{Lynx2020} & \cite{RaDD2020} & \cite{SmarterNIC2022} & \cite{MolecularDynamics2022} & \cite{derda2023SCposter} & \cite{DNN2022} & \cite{HeterogeneousSmartNIC01} & \cite{LightningZhong} \\
    \hline 
    Application \\
        \hspace{5mm} Broad Topic                                     & OpenMP    & MPI       & Infrastr. & Infrastr. & Simulation & Simulation & GIS     & AI/ML     & AI/ML     & Optical   \\
        \hspace{5mm} Exploration of non-infrastructure application   & $\bullet$ & $\bullet$ & $       $ & $\CIRCLE$ & $\CIRCLE$ & $\CIRCLE$ & $\CIRCLE$ & $\CIRCLE$ & $\CIRCLE$ & $\CIRCLE$ \\
        \hspace{5mm} Specialized use beyond just as a CPU node       & $       $ & $       $ & $\CIRCLE$ & $       $ & $\bullet$ & $\CIRCLE$ & $       $ & $       $ & $\CIRCLE$ & $\CIRCLE$ \\
    \hline 
    Hardware and Device Usage \\
        \hspace{5mm} Involves DPU (or exotic SmartNIC)               & $\CIRCLE$ & $\CIRCLE$ & $\CIRCLE$ & $\CIRCLE$ & $\CIRCLE$ & $\CIRCLE$ & $\CIRCLE$ & $\CIRCLE$ & $\CIRCLE$ & $\bullet$ \\
        \hspace{5mm} Use of SmartNIC emulation                       & $       $ & $       $ & $       $ & $       $ & $       $ & $       $ & $       $ & $       $ & $       $ & $\CIRCLE$ \\
        \hspace{5mm} Use of SmartNIC ASICs                           & $       $ & $       $ & $       $ & $       $ & $       $ & $\CIRCLE$ & $       $ & $       $ & $       $ & $\bullet$ \\
        \hspace{5mm} Use of FPGAs or custom hardware                 & $       $ & $       $ & $\CIRCLE$ & $       $ & $       $ & $\bullet$ & $       $ & $       $ & $\CIRCLE$ & $\CIRCLE$ \\
    \hline 
    Contribution \\
        \hspace{5mm} Presents an implementation or experimentation   & $\CIRCLE$ & $\CIRCLE$ & $\CIRCLE$ & $       $ & $\CIRCLE$ & $       $ & $\CIRCLE$ & $\CIRCLE$ & $\CIRCLE$ & $\bullet$ \\
        \hspace{5mm} Includes in-depth benchmark(s)                  & $\bullet$ & $\CIRCLE$ & $\CIRCLE$ & $       $ & $\CIRCLE$ & $       $ & $\bullet$ & $\bullet$ & $\CIRCLE$ & $\bullet$ \\
        \hspace{5mm} Speedup/throughput improvement (rounded)        & $-      $ & $2\times$ & $4\times$ & $-      $ & $1\times$ & $-      $ & $2\times$ & $1\times$ & $1\times$ & $300\times$ \\
        \hspace{5mm} Library or framework contribution               & $\CIRCLE$ & $\CIRCLE$ & $\CIRCLE$ & $       $ & $       $ & $\bullet$ & $       $ & $       $ & $       $ & $\bullet$ \\
        \hspace{5mm} Algorithmic contribution                        & $       $ & $\CIRCLE$ & $       $ & $       $ & $\CIRCLE$ & $\bullet$ & $       $ & $       $ & $\CIRCLE$ & $\bullet$ \\
        \hspace{5mm} Page count + abstract                           & $8      $ & $6, 20  $ & $15     $ & $8      $ & $12     $ & $7      $ & $3      $ & $8      $ & $12     $ & $21     $ \\
    \hline 
    Discussion \\
        \hspace{5mm} Discusses scalability                           & $       $ & $\bullet$ & $\CIRCLE$ & $       $ & $       $ & $\bullet$ & $\CIRCLE$ & $\CIRCLE$ & $\CIRCLE$ & $       $ \\
        \hspace{5mm} Identifies further research gaps                & $\bullet$ & $\bullet$ & $       $ & $\CIRCLE$ & $\CIRCLE$ & $\CIRCLE$ & $\bullet$ & $       $ & $       $ & $       $ \\
        \hspace{5mm} Discusses limitations of own work               & $       $ & $       $ & $       $ & $       $ & $\CIRCLE$ & $       $ & $\bullet$ & $       $ & $       $ & $       $ \\
    \hline 
    Parallelism \\
        \hspace{5mm} Exhibits shared-memory / thread parallelism     & $\CIRCLE$ & $       $ & $\bullet$ & $\bullet$ & $\CIRCLE$ & $\bullet$ & $       $ & $       $ & $       $ & $       $ \\
        \hspace{5mm} Exhibits distributed memory parallelism         & $       $ & $\CIRCLE$ & $\CIRCLE$ & $\CIRCLE$ & $\CIRCLE$ & $\CIRCLE$ & $\CIRCLE$ & $\CIRCLE$ & $\CIRCLE$ & $\bullet$ \\
        \hspace{5mm} Uses multiple SmartNIC nodes                    & $\CIRCLE$ & $\CIRCLE$ & $       $ & $       $ & $\CIRCLE$ & $\CIRCLE$ & $\CIRCLE$ & $\CIRCLE$ & $\CIRCLE$ & $       $ \\
        \hspace{5mm} Exhibits novel pipeline parallelism             & $       $ & $\bullet$ & $       $ & $       $ & $\CIRCLE$ & $       $ & $       $ & $       $ & $\bullet$ & $\CIRCLE$ \\
        \hspace{5mm} Exhibits use of internal SIMD capabilities      & $\CIRCLE$ & $       $ & $       $ & $       $ & $\CIRCLE$ & $       $ & $       $ & $       $ & $       $ & $\CIRCLE$ \\
        \hspace{5mm} Uses external SIMD (e.g. GPU) devices           & $\CIRCLE$ & $       $ & $\CIRCLE$ & $\bullet$ & $       $ & $       $ & $       $ & $       $ & $\CIRCLE$ & $       $ \\
    \hline 
    
\end{tabular}
}
\caption{
    An analysis of various traits of the papers most closely concerned with parallel and distributed use of SmartNICs, identifying whether each work \textbf{strongly}~($\CIRCLE$), \textbf{weakly}~($\bullet$), or \textbf{does not} clearly appear to possess the trait.
    Note that two related MPI papers have been treated as one unit (second column).
}

\label{table:parallel}
\end{table*}



%% file: Table_5.tex
\begin{table*}[h]
\small
\centering
\begin{tabular}{lll} 
\hline
    \textbf{Topic or Benchmark Category} & \textbf{General}                                                       & \textbf{Contribution-Specific} \\ \hline
    Works Creating Benchmark Suites      & \cite{hu2025dpbento, kashyap_idiosyncrasies, dpubench2023, wang2023dpubench}                   & \\
    Non-Benchmark Studies                & \cite{comprehensive_expert_survey, in_network_computing_survey, smartnics2021, Caulfield2018, RaDD2020, dpuedge2022, DDoS2019, sherry2024driven,
                                         dpubasedhardwareacceleration, gRPC2021}                                  & \\
    Whitepapers \& Specs                 & \cite{increasingdatacenter, nvidiaArch, bf2datasheet_2023, awsnitrowhitepaper, liquid3, dpubasedhardwareacceleration, Intelipu, amdpensando2023, converged_accelerators, whatisadpu, whatisasupernic} & \\ \hline
    Compute Power                        & \cite{ComprehensiveStudyDPU2023, PerformanceCharacteristics2021, hu2025dpbento, smartNICsiPipe2019, chen2024datapathaccelerator} & \cite{styx2023, suresh_krylov, huang_DORM} \\
    Communication (Throughput, Latency)  & \cite{comprehensive_expert_survey, ComprehensiveStudyDPU2023, CharacterizingSmartNIC2023, PerformanceCharacteristics2021, kashyap_idiosyncrasies, hu2025dpbento, smartNICsiPipe2019, SmarterNIC2022, nvidiaArch}
                                                                                                                  & \cite{xenic2021, SmarterNIC2022, Xing2023, Taehyun2023, latency01, Large-Message_MPI, styx2023, FairNIC2020, SuperNIC, Li2022, PsPIN2020} \\
    Memory Access (Local, Distributed)   & \cite{ComprehensiveStudyDPU2023, PerformanceCharacteristics2021, kashyap_idiosyncrasies, hu2025dpbento, smartNICsiPipe2019} & \cite{Zhang2023} \\
    Storage (Local, Distributed)         & \cite{PerformanceCharacteristics2021, kashyap_idiosyncrasies, hu2025dpbento, nvidiaArch}       & \cite{CharacterizingSmartNIC2023, DPC01, inec2020,                                                                                                      D2Comp2024, DComp2023, Kim2021, xenic2021, sPIN:High-performance, PsPIN2020} \\
    Compression                          & \cite{comprehensive_expert_survey, ProceedingswithsmartNIC, LossyandLosslessCompression02, PerformanceCharacteristics2021, hu2025dpbento}
                                                                                                                  & \cite{AcceleratingLossy01, Runway2023, D2Comp2024, DComp2023}  \\
    Other ASIC Accelerators              & \cite{comprehensive_expert_survey, ComprehensiveStudyDPU2023, PerformanceCharacteristics2021, hu2025dpbento, chen2024datapathaccelerator}       & \\ \hline
    Applications, System Services        & \cite{nvidiaArch}                                                      & \cite{HeterogeneousSmartNIC01, SmarterNIC2022, FPGA2022, smartNICsiPipe2019, Lynx2020,                                                                                                  Lin2023, inventifationMLBluefieldDPU, derda2023SCposter, tong_DALdex, suresh_krylov, huang_DORM} \\ 
    Engergy Savings                      & \cite{increasingdatacenter}                                            & \cite{DPUweapondetectioncase2022, LightningZhong, zawawi2023resource, EdgeForElasticity} \\
    Scalability                          &                                                                        & \cite{SmarterNIC2022, dpunumerousvpnconnection, HeterogeneousSmartNIC01, FCsN_NN_FPGA,                                                                                                                              Siracusano2022, DNN2022} \\ \hline

\end{tabular}
\caption{Works performing noteworthy studies, or benchmarks of DPU hardware, organized by benchmarking criteria. Works in the far right column simply measure the effects of their own specific contribution.}
\label{table:benchmarks}
\end{table*}

%% file: Table_6.tex
\begin{table}[h]
\centering
\resizebox{\linewidth}{!}{
\begin{tabular}{lp{155pt}} \hline 
    \textbf{Subdomain or Offload}       & \textbf{References}                                                                                                                                 \\ \hline 
                                                                                                                                                                                                       
    Cloud Architecture                  & \cite{Caulfield2018, in_network_computing_survey, Lynx2020, dpuedge2022, sherry2024driven, increasingdatacenter, awsnitrowhitepaper, liquid3, RaDD2020, nvidiaArch, gRPC2021} 
                           \\ \hline
    Network Communication               & \cite{Li2022, NovelFramework2023ieee, styx2023, SmarterNIC2022, MolecularDynamics2022, xenic2021, Lynx2020, PsPIN2020, dpubench2023,                                                                smartNICsiPipe2019, PerformanceCharacteristics2021, HeterogeneousSmartNIC01, FCsN_NN_FPGA, Siracusano2022, latency01, sPIN:High-performance, CIDR25} \\
    \hspace{5mm} Message Proc., TCP/IP  & \cite{Taehyun2023, Zhang2023, huang_DORM}                                                                                           \\
    \hspace{5mm} Pipelines, P4          & \cite{Xing2023, p4smartnic2020, SuperNIC}                                                                                                           \\
    \hspace{5mm} Content Delivery       & \cite{Taehyun2023}                                                                                                                                  \\ \hline
    Hardware Implementation, FPGAs      & \cite{Zhang2023, Li2022, HyperionAC2022, SuperNIC} (and Section \ref{section_ai_ml})                                               \\ \hline
    Distributed Storage                 & \cite{Zhang2023, dpubench2023, dpfs2023, CIDR25}                                                                                                                      \\
    \hspace{5mm} File Systems           & \cite{Kim2021, CharacterizingSmartNIC2023, DPC01}                                                                                                   \\
    \hspace{5mm} Compression            & \cite{Zhang2023, D2Comp2024, DComp2023, AcceleratingLossy01, LossyandLosslessCompression02, ProceedingswithsmartNIC, styx2023,                                                   Runway2023} \\
    \hspace{5mm} Databases \& KV-Stores & \cite{CharacterizingSmartNIC2023, D2Comp2024, DComp2023, inec2020, FairNIC2020, EdgeForElasticity, inventifationMLBluefieldDPU}                     \\ \hline
    Big-Data Processes                  & \cite{Lin2023, smartNICsiPipe2019}                                                                                                                  \\ \hline
    5G Networks                         & \cite{p4smartnic2020, FPGA2022, dpuedge2022}                                                                                                        \\ \hline
    Virtualization, Containers          & \cite{huang_DORM, FairNIC2020, uno2017, SuperNIC, dpunumerousvpnconnection, dpusolution2022}                                                        \\ \hline
    Security                            & \cite{dpunumerousvpnconnection, dpuedge2022, DDoS2019, dpubench2023, awsnitrowhitepaper}                                                            \\ \hline
    Kernel                              & \cite{styx2023}                                                                                                                                     \\ \hline
    Blockchain                          & \cite{BlockNIC2023, BlockchainSmartNIC2023}                                                                                                         \\ \hline
    Scientific HPC (Section \ref{section_HPC}) & \cite{SmarterNIC2022, MolecularDynamics2022, ProceedingswithsmartNIC, composableDataServices2023, dpubench2023, Runway2023, AcceleratingLossy01, DNN2022,                                               derda2023SCposter, tong_DALdex, suresh_krylov} \\ \hline 
    AI, ML                              & (See Table \ref{table:neural_net_type})                                                                                                             \\ \hline
    
    \hline
\end{tabular}
}
\caption{Works using SmartNICs in various commercial applications and network or data center infrastructure, organized by subdomain or topic. Works in each of our tables may appear in multiple categories.}
\label{table:infr_and_app}
\end{table}


%% file: Table_7.tex
\begin{table}[h]
\centering
\small
\begin{tabular}{ll} \hline 
    \textbf{AI/ML Topic}                & \textbf{References}                         \\ \hline
           
    DPUs + AI / ML                      & \cite{dpubasedhardwareacceleration, EdgeForElasticity, FCsN_NN_FPGA, inventifationMLBluefieldDPU} \\
    \hspace{5mm} Deep Learning          & \cite{LightningZhong, DNN2022, styx2023} \\
    \hspace{5mm} Binary NN              & \cite{Siracusano2022} \\
    \hspace{5mm} Graph NN               & \cite{FCsN_NN_FPGA} \\
    \hspace{5mm} Reinforcement Learning & \cite{latency01} \\
    \hspace{5mm} Federated Learning     & \cite{federatedlearning} \\
    \hspace{5mm} Computer Vision        & \cite{DPUweapondetectioncase2022, EdgeForElasticity} \\ \hline
    In-Hardware / FPGA Impl.            & \cite{LightningZhong, FCsN_NN_FPGA, latency01} \\ \hline 
    DPUs + GPUs                         & \cite{Lynx2020, dpubasedhardwareacceleration, Runway2023, HeterogeneousSmartNIC01, zawawi2023resource} \\ \hline
    
    \hline
\end{tabular}
\caption{Works relating to using DPUs for AI \& ML, organized by subtopic or neural network type.}
\label{table:neural_net_type}
\end{table}